\documentclass[english,superscriptaddress]{revtex4-1}
\usepackage[T1]{fontenc}
\usepackage[latin9]{inputenc}
\usepackage{color}
\usepackage{babel}
\usepackage{verbatim}
\usepackage{amsmath}
\usepackage{amssymb}
\usepackage{graphicx}
\usepackage{wasysym}
\usepackage[unicode=true,pdfusetitle,
 bookmarks=true,bookmarksnumbered=false,bookmarksopen=false,
 breaklinks=false,pdfborder={0 0 1},backref=false,colorlinks=true]
 {hyperref}

\makeatletter
\@ifundefined{showcaptionsetup}{}{%
 \PassOptionsToPackage{caption=false}{subfig}}
\usepackage{subfig}
\makeatother

\begin{document}
\title{Spin density matrix for neutral $\rho$ mesons in a pion gas in linear
response theory}
\author{Yi-Liang Yin}
\affiliation{Department of Modern Physics and Anhui Center for Fundamental Sciences
in Theoretical Physics, University of Science and Technology of China,
Hefei, Anhui 230026, China}
\author{Wen-Bo Dong }
\affiliation{Department of Modern Physics and Anhui Center for Fundamental Sciences
in Theoretical Physics, University of Science and Technology of China,
Hefei, Anhui 230026, China}
\author{Cong Yi}
\affiliation{Department of Modern Physics and Anhui Center for Fundamental Sciences
in Theoretical Physics, University of Science and Technology of China,
Hefei, Anhui 230026, China}
\author{Qun Wang}
\affiliation{Department of Modern Physics and Anhui Center for Fundamental Sciences
in Theoretical Physics, University of Science and Technology of China,
Hefei, Anhui 230026, China}
\affiliation{School of Mechanics and Physics, Anhui University of Science and Technology,
Huainan, Anhui 232001, China}
\begin{abstract}
We calculate the spin density matrix for neutral $\rho$ mesons from
the spectral function and thermal shear tensor by Kubo formula in
the linear response theory, which contributes to the $\gamma$ correlator
for the CME search. We derive the spectral function of neutral $\rho$
mesons with $\rho\pi\pi$ and $\rho\rho\pi\pi$ interactions using
the Dyson-Schwinger equation. The thermal shear tensor contribution
is obtained from the Kubo formula in the linear response theory. We
numerically calculate $\rho_{00}-1/3$ and $\mathrm{Re}\rho_{-1,1}$
using the simulation results for the thermal shear tensor by the hydrodynamical
model, which are of the order $10^{-3}\sim10^{-2}$.
\end{abstract}
\maketitle
%done, 2025.2.6, 11:30

\section{Introduction}

The spin and orbital angular momentum are intrinsically connected
and can be converted to each other as demonstrated in the Barnett
effect \citep{Barnett:1935} and Einstein-de-Haas effect \citep{dehaas:1915}
in materials. In non-central heavy-ion collisions, a huge orbital
angular momentum is generated and partially transferred to the strongly
interacting matter in the form of particle's spin polarization along
the reaction plane through spin-orbit couplings \citep{Liang:2004ph}.
This phenomenon is called the global spin polarization as it is with
respect to the reaction plane which is fixed for all particles in
one event \citep{Liang:2004ph,Voloshin:2004ha,Liang:2004xn,Betz:2007kg,Gao:2007bc,Becattini:2007sr}.
The global spin polarization of $\Lambda$ and $\overline{\Lambda}$
hyperons was first observed and measured in the STAR experiment \citep{STAR:2017ckg,STAR:2018gyt},
and then measured in experiments of HADES \citep{HADES:2022enx} and
ALICE \citep{ALICE:2021pzu}. Since its first observation \citep{STAR:2017ckg,STAR:2018gyt},
there has been tremendous advance in the study of the global spin
polarization in theoretical models \citep{Karpenko:2016jyx,Li:2017slc,Xie:2017upb,Sun:2017xhx,Baznat:2017jfj,Shi:2017wpk,Wei:2018zfb,Fu:2020oxj,Ryu:2021lnx,Deng:2021miw,Wu:2022mkr},
see Refs. \citep{Wang:2017jpl,Florkowski:2018fap,Huang:2020dtn,Gao:2020lxh,Gao:2020vbh,Liu:2020ymh,Becattini:2022zvf,Hidaka:2022dmn,Becattini:2024uha}
for recent reviews on this topic.

%done, 2025.2.6, 5:40

The spin polarization of hyperons can be measured through their parity-breaking
weak decays \citep{Bunce:1976yb}. Vector mesons mostly decay through
parity-conserved strong interaction, so there is no way to measure
the spin polarization of vector mesons. The spin states of the vector
meson can be described by the spin density matrix $\rho_{\lambda_{1}\lambda_{2}}$,
where $\lambda_{1}$ and $\lambda_{2}=0,\pm1$ denote the spin states.
The spin density matrix is normalized $\mathrm{tr}\rho\equiv\sum_{\lambda}\rho_{\lambda\lambda}=1$.
The $\rho_{00}$ element can be measured in experiments through the
polar angle distribution of decay products \citep{Schilling:1969um,Liang:2004xn,Yang:2017sdk,Tang:2018qtu}.
If $\rho_{00}\neq1/3$, the spin-0 state of the vector meson is not
equally populated as $\lambda=\pm1$ states, which is called the spin
alignment. Specifically, $\rho_{00}<1/3$ means the field vector (not
the spin vector) is more aligned to the direction perpendicular to
the spin quantization direction, while $\rho_{00}>1/3$ means the
field vector is more aligned to the spin quantization direction. The
spin alignment with respect to the reaction plane is called the global
spin alignment and was also predicted in 2004-2005 \citep{Liang:2004xn}.
The global spin alignment of $\phi$ and $K^{0*}$ mesons has recently
been measured by the STAR collaboration in Au+Au collisions \citep{STAR:2022fan},
where a large deviation of $\rho_{00}$ for $\phi$ mesons from 1/3
is observed at lower collision energies, but there is no significant
deviation of $\rho_{00}$ for $K^{0*}$ mesons within errors.

%done, 2025.2.6, 6:00

A number of sources can contribute to the spin alignment of $\phi$
mesons but not enough to explain such a large effect \citep{Yang:2017sdk,Xia:2020tyd,Gao:2021rom,Muller:2021hpe,Kumar:2023ghs}.
It was proposed that a large spin alignment of $\phi$ mesons may
come from the local correlation/fluctuation of strong vector fields
that are coupled to s and $\bar{\mathrm{s}}$ quarks \citep{Sheng:2019kmk}.
Initially a nonrelativistic quark coalescence model \citep{Yang:2017sdk}
was used to calculate the spin density matrix of $\phi$ mesons with
s and $\bar{\mathrm{s}}$ quarks being polarized by strong vector
fields \citep{Sheng:2020ghv}. Then the nonrelativistic quark coalescence
model was generalized to a relativistic one \citep{Sheng:2022ffb,Sheng:2022wsy}.
The relativistic model is based on Wigner functions and spin kinetic
equations for vector mesons. The model incorporated with strong vector
fields can successfully describe the experimental data for the global
spin alignment of $\phi$ mesons \citep{Sheng:2022ffb,Sheng:2022wsy,Sheng:2023urn}.
For recent reviews of the topic, the readers may look at Refs. \citep{Chen:2023hnb,Sheng:2023chinphyb,Chen:2024bik,Chen:2024afy}.

%done, 2025.2.6, 7:20

Recently, the thermal shear contribution to $\phi$ meson's spin alignment
has been studied in the linear response theory \citep{Li:2022vmb,Dong:2024nxj}.
In this paper, we will apply the same method to another vector meson,
the neutral rho meson or $\rho^{0}$. The spin density matrix of $\rho^{0}$
may serve as a background for the chiral magnetic effect (CME) \citep{Kharzeev:2004ey,Kharzeev:2007jp,Fukushima:2008xe},
because the anisotropic angular distribution of its decay daughters
(pions) provides a non-vanishing contribution to the $\gamma$ correlator
\citep{Voloshin:2004vk,STAR:2013ksd,STAR:2013zgu,Wang:2016iov}, see
Refs. \citep{Tang:2019pbl,Shen:2022gtl} for quantitative analyses.
In Ref. \citep{Yin:2024dnu} by some of us, the evolution of $\rho^{0}$
mesons in a pion gas was studied by the spin kinetic or Boltzmann
equation for the on-shell spin dependent distribution. The results
show that the spin alignment of $\rho^{0}$ mesons is damped rapidly
in the pion gas due to $\rho\pi\pi$ interaction and that $\rho_{00}$
and $\mathrm{Re}\rho_{-1,1}$ have a significant contribution to the
$\gamma$ correlator for CME. In this paper, we will consider the
off-shell contribution as well as the thermal shear contribution to
$\rho_{00}$ and $\mathrm{Re}\rho_{-1,1}$ by assuming thermal spin
distributions for $\rho^{0}$. We will treat the thermal shear tensor
as a perturbation from local equilibrium and apply the Kubo formula
\citep{Zubarev_1979,Hosoya:1983id,Becattini:2019dxo} to calculate
the linear response of the spin density matrix for $\rho^{0}$ mesons
as done for $\phi$ mesons in Refs. \citep{Li:2022vmb,Dong:2024nxj}.
The spectral function of $\rho^{0}$ mesons is derived from $\rho\pi\pi$
and $\rho\rho\pi\pi$ interaction in the chiral effective theory \citep{Gale:1990pn}.
Finally, we present numerical results for $\rho_{00}$ and $\mathrm{Re}\rho_{-1,1}$
of $\rho^{0}$ mesons using the simulation results for the thermal
shear tensor from the hydrodynamical model \citep{Pang:2018zzo,Wu:2021fjf,Wu:2022mkr}.

%done, 2025.2.6, 9:40

The paper is organized as follows. In Sec. \ref{sec:CTP-Green's-function},
we briefly introduce the Green's functions for vector mesons and pseudoscalar
mesons in closed-time-path (CTP) formalism. In Sec. (\ref{sec:Wigner-transformation-and}),
we define the Wigner function and the matrix valued spin dependent
distribution (MVSD) for the $\rho^{0}$ meson which is proportional
to its spin density matrix. In Sec. (\ref{sec:Dyson-Schwinger-Equation-and}),
we use the Dyson-Schwinger equation to derive the spectral function
of the $\rho^{0}$ meson. We also derive the explicit form of the
Wigner function without nonlocal terms. In Sec. (\ref{sec:Linear-Response-Theory}),
we obtain the nonlocal correction to the Wigner function proportional
to the thermal shear tensor through the Kubo formula in the linear
response theory. The numerical results are presented in Sec. (\ref{sec:Numerical-Result})
and a summary of results and the conclusion are given in the last
section.

%done, 2025.2.6, 10:40

By convention, the metric tensor is $g^{\mu\nu}=\mathrm{diag}(1,-1,-1,-1)$,
four-momenta are denoted as $p^{\mu}=(p_{0},\mathbf{p})$, $p_{\mu}=(p_{0},-\mathbf{p})$,
where $\mu,\nu=$0,1,2,3, and Greek letters and Latin letters represent
space-time components of four-vectors and space components of three-vectors
respectively. For notational clarification, we use $G^{\mu\nu}$ and
$p$ for the two-point Green's function and momentum for the $\rho^{0}$
meson, while $S$ and $k$ for the two-point Green's function and
momentum for the pseudoscalar meson $\pi^{\pm}$.

%done, 2025.2.6, 10:40

\section{Green's function in closed-time-path formalism \label{sec:CTP-Green's-function} }

The closed-time-path (CTP) or Schwinger-Keldysh formalism in quantum
field theory is an effective method to describe equilibrium and non-equilibrium
physics in many-body systems \citep{Martin:1959jp,Keldysh:1964ud,Chou:1984es,Blaizot:2001nr,Wang:2001dm,Berges:2004yj,Cassing:2008nn,Crossley:2015evo}.
The two-point Green's functions on the CTP for the spin-1 vector meson
($\rho^{0}$) and spin-0 pseudoscalar meson $(\pi^{\pm})$ are defined
as
\begin{eqnarray}
G_{\mathrm{CTP}}^{\mu\nu}(x_{1},x_{2}) & = & \left\langle T_{C}A^{\mu}(x_{1})A^{\nu}(x_{2})\right\rangle ,\nonumber \\
S_{\mathrm{CTP}}(x_{1},x_{2}) & = & \left\langle T_{C}\phi(x_{1})\phi^{\dagger}(x_{2})\right\rangle ,\label{eq:green-function}
\end{eqnarray}
where $T_{C}$ denotes the time-ordered operator on the CTP, and $A^{\mu}$
and $\phi$ are the real vector field and complex scalar field respectively,
which can be quantized as
\begin{eqnarray}
A^{\mu}(x) & = & \sum_{\lambda=0,\pm1}\int\frac{d^{3}p}{(2\pi)^{3}}\frac{1}{2E_{p}^{\rho}}\nonumber \\
 &  & \times\left[\epsilon^{\mu}(\lambda,{\bf p})a_{V}(\lambda,{\bf p})e^{-ip\cdot x}+\epsilon^{\mu\ast}(\lambda,{\bf p})a_{V}^{\dagger}(\lambda,{\bf p})e^{ip\cdot x}\right],\nonumber \\
\phi(x) & = & \int\frac{d^{3}k}{(2\pi)^{3}}\frac{1}{2E_{k}^{\pi}}\left[a({\bf k})e^{-ik\cdot x}+b^{\dagger}({\bf k})e^{ik\cdot x}\right],\label{eq:quant-field}
\end{eqnarray}
where $p=(E_{p}^{\rho},\mathbf{p})$ and $k=(E_{k}^{\pi},\mathbf{k})$
are on-shell momenta for $\rho$ and $\pi$ with $E_{p}^{\rho}=\sqrt{\mathbf{p}^{2}+m_{\rho}^{2}}$
and $E_{k}^{\pi}=\sqrt{\mathbf{k}^{2}+m_{\pi}^{2}}$, $\lambda$ denotes
the spin state, and $\epsilon^{\mu}(\lambda,{\bf p})$ is the spin
polarization vector
\begin{eqnarray}
\epsilon^{\mu}(\lambda,\mathbf{p}) & = & \left(\frac{\mathbf{p}\cdot\boldsymbol{\epsilon}_{\lambda}}{m_{\rho}},\boldsymbol{\epsilon}_{\lambda}+\frac{\mathbf{p}\cdot\boldsymbol{\epsilon}_{\lambda}}{m_{\rho}(E_{p}^{\rho}+m_{\rho})}\mathbf{p}\right),
\end{eqnarray}
where $\boldsymbol{\epsilon}_{\lambda}$ is the spin polarization
three-vector in the particle's rest frame given by
\begin{align}
\boldsymbol{\epsilon}_{0}= & (0,1,0),\nonumber \\
\boldsymbol{\epsilon}_{+1}= & -\frac{1}{\sqrt{2}}(i,0,1),\nonumber \\
\boldsymbol{\epsilon}_{-1}= & \frac{1}{\sqrt{2}}(-i,0,1).
\end{align}
We have chosen $+y$ to be the spin quantization direction. The polarization
vector satisfies
\begin{eqnarray}
p_{\mu}\epsilon^{\mu}(\lambda,{\bf p}) & = & 0,\nonumber \\
\epsilon(\lambda,{\bf p})\cdot\epsilon^{*}(\lambda^{\prime},{\bf p}) & = & -\delta_{\lambda\lambda^{\prime}},\nonumber \\
-\sum_{\lambda}\epsilon^{\mu}(\lambda,{\bf p})\epsilon^{\nu*}(\lambda,{\bf p}) & = & \Delta^{\mu\nu}(p_{\mathrm{on}})\equiv g^{\mu\nu}-\frac{p^{\mu}p^{\nu}}{m_{\rho}^{2}}.\label{eq:epsilon-property}
\end{eqnarray}
Note that in the above equation, $p^{\mu}$ is an on-shell momentum,
so we put the index ``on'' to the momentum variable in the projector
$\Delta^{\mu\nu}(p_{\mathrm{on}})$ to distinguish it from the off-shell
projector $\Delta^{\mu\nu}(p)$ which will be used later. We see in
Eq. (\ref{eq:quant-field}) that we use $G$ and $p$ to denote the
Green's function and momentum for $\rho^{0}$ respectively, while
we use $S$ and $k$ to denote the Green's function and momentum for
$\pi^{\pm}$ respectively.

%done, 2025.2.3, 7:00

In the CTP formalism, $G_{\mathrm{CTP}}^{\mu\nu}$ in Eq. (\ref{eq:green-function})
has four components
\begin{eqnarray}
G_{F}^{\mu\nu}(x_{1},x_{2}) & \equiv & \theta(t_{1}-t_{2})\left\langle A^{\mu}(x_{1})A^{\nu}(x_{2})\right\rangle +\theta(t_{2}-t_{1})\left\langle A^{\nu}(x_{2})A^{\mu}(x_{1})\right\rangle ,\nonumber \\
G_{<}^{\mu\nu}(x_{1},x_{2}) & \equiv & \left\langle A^{\nu}(x_{2})A^{\mu}(x_{1})\right\rangle ,\nonumber \\
G_{>}^{\mu\nu}(x_{1},x_{2}) & \equiv & \left\langle A^{\mu}(x_{1})A^{\nu}(x_{2})\right\rangle ,\nonumber \\
G_{\overline{F}}^{\mu\nu}(x_{1},x_{2}) & \equiv & \theta(t_{2}-t_{1})\left\langle A^{\mu}(x_{1})A^{\nu}(x_{2})\right\rangle +\theta(t_{1}-t_{2})\left\langle A^{\nu}(x_{2})A^{\mu}(x_{1})\right\rangle ,\label{eq:Green-function-less}
\end{eqnarray}
depending on whether $x_{1}$ and $x_{2}$ are on the positive or
negative time branch. One can verify that all four components satisfy
the identity $G_{F}^{\mu\nu}+G_{\overline{F}}^{\mu\nu}=G_{<}^{\mu\nu}+G_{>}^{\mu\nu}$,
so only three of them are independent which one can choose the retarded,
advanced and correlation Green's functions $G_{R}^{\mu\nu}$, $G_{A}^{\mu\nu}$
and $G_{C}^{\mu\nu}$ as
\begin{eqnarray}
G_{R}^{\mu\nu}(x_{1},x_{2}) & \equiv & \theta(t_{1}-t_{2})\left(\left\langle A^{\mu}(x_{1})A^{\nu}(x_{2})\right\rangle -\left\langle A^{\nu}(x_{2})A^{\mu}(x_{1})\right\rangle \right)\nonumber \\
 & = & G_{F}^{\mu\nu}(x_{1},x_{2})-G_{<}^{\mu\nu}(x_{1},x_{2}),\nonumber \\
G_{A}^{\mu\nu}(x_{1},x_{2}) & \equiv & \theta(t_{2}-t_{1})\left(\left\langle A^{\nu}(x_{2})A^{\mu}(x_{1})\right\rangle -\left\langle A^{\mu}(x_{1})A^{\nu}(x_{2})\right\rangle \right)\nonumber \\
 & = & G_{F}^{\mu\nu}(x_{1},x_{2})-G_{>}^{\mu\nu}(x_{1},x_{2}),\nonumber \\
G_{C}^{\mu\nu}(x_{1},x_{2}) & \equiv & \left\langle A^{\mu}(x_{1})A^{\nu}(x_{2})\right\rangle +\left\langle A^{\nu}(x_{2})A^{\mu}(x_{1})\right\rangle \nonumber \\
 & = & G_{<}^{\mu\nu}(x_{1},x_{2})+G_{>}^{\mu\nu}(x_{1},x_{2}).\label{eq:Green-function-R}
\end{eqnarray}
The four components of the Green's function for the complex scalar
field can be defined similarly. Other operators can also be defined
on the CTP, obeying the relations similar to Eq. (\ref{eq:Green-function-R})
as
\begin{eqnarray}
O_{R}(x_{1},x_{2}) & = & O_{F}(x_{1},x_{2})-O_{<}(x_{1},x_{2}),\nonumber \\
O_{A}(x_{1},x_{2}) & = & O_{F}(x_{1},x_{2})-O_{>}(x_{1},x_{2}),\nonumber \\
O_{C}(x_{1},x_{2}) & = & O_{<}(x_{1},x_{2})+O_{>}(x_{1},x_{2}),
\end{eqnarray}
where $O$ is any operator defined on the CTP.

%done, 2025.2.3, 7:20

\section{Wigner functions and spin density matrix \label{sec:Wigner-transformation-and}}

In order to define the matrix valued spin dependent distribution (MVSD)
for the vector meson in phase space, we introduce the Wigner transformation
for the two-point Green's function that defines the Wigner function
\begin{eqnarray}
G_{<}^{\mu\nu}(x,p) & = & \int d^{4}ye^{ip\cdot y}G_{<}^{\mu\nu}\left(x+\frac{1}{2}y,x-\frac{1}{2}y\right).\label{eq:Wigner-tans}
\end{eqnarray}
For free particles, the Wigner function has the form
\begin{eqnarray}
G_{(0)<}^{\mu\nu}(x,p) & = & 2\pi\sum_{\lambda_{1},\lambda_{2}}\delta\left(p^{2}-m_{V}^{2}\right)\left\{ \theta(p^{0})\epsilon^{\mu}\left(\lambda_{1},{\bf p}\right)\epsilon^{\nu\ast}\left(\lambda_{2},{\bf p}\right)f_{\lambda_{1}\lambda_{2}}^{(0)}(x,{\bf p})\right.\nonumber \\
 &  & \left.+\theta(-p^{0})\epsilon^{\mu\ast}\left(\lambda_{1},-{\bf p}\right)\epsilon^{\nu}\left(\lambda_{2},-{\bf p}\right)\left[\delta_{\lambda_{2}\lambda_{1}}+f_{\lambda_{2}\lambda_{1}}^{(0)}(x,-{\bf p})\right]\right\} ,\label{eq:Gless0}
\end{eqnarray}
where the index ``(0)'' denotes the leading order of $\hbar$, and
$f_{\lambda_{1}\lambda_{2}}^{(0)}(x,{\bf p})$ is the MVSD at leading
order defined as
\begin{eqnarray}
f_{\lambda_{1}\lambda_{2}}^{(0)}(x,{\bf p}) & \equiv & \int\frac{d^{4}u}{2(2\pi)^{3}}\delta(p\cdot u)e^{-iu\cdot x}\left\langle a_{V}^{\dagger}\left(\lambda_{2},{\bf p}-\frac{{\bf u}}{2}\right)a_{V}\left(\lambda_{1},{\bf p}+\frac{{\bf u}}{2}\right)\right\rangle ,\label{eq:MVSD}
\end{eqnarray}
which satisfies $f_{\lambda_{1}\lambda_{2}}^{(0)*}(x,{\bf p})=f_{\lambda_{2}\lambda_{1}}^{(0)}(x,{\bf p})$.
Note that the MVSD can be decomposed into the scalar (trace), polarization
($P_{i}$) and tensor ($T_{ij}$) parts as \citep{Sheng:2023chinphyb,Becattini:2024uha}
\begin{equation}
f_{\lambda_{1}\lambda_{2}}^{(0)}(x,{\bf p})=\mathrm{Tr}(f^{(0)})\left(\frac{1}{3}+\frac{1}{2}P_{i}\Sigma_{i}+T_{ij}\Sigma_{ij}\right)_{\lambda_{1}\lambda_{2}},\label{eq:decom-f}
\end{equation}
where $i,j=1,2,3$, $\mathrm{Tr}(f^{(0)})=\sum_{\lambda}f_{\lambda\lambda}^{(0)}$,
and $\Sigma_{i}$ and $\Sigma_{ij}$ are traceless matrices defined
in Eq. (11) of Ref. \citep{Dong:2023cng}. The on-shell part of the
Wigner function $W^{\mu\nu}(x,\mathbf{p})$ can be obtained by integration
of $(p_{0}/\pi)G_{(0)<}^{\mu\nu}(x,p)$ over $p_{0}=\left[0,\infty\right)$.
Using the decomposition (\ref{eq:decom-f}) $W^{\mu\nu}(x,\mathbf{p})$
can be decomposed into the scalar ($\mathcal{S}$), polarization ($G^{[\mu\nu]}$)
and tensor ($\mathcal{T}^{\mu\nu}$) parts as
\begin{equation}
W^{\mu\nu}(x,\mathbf{p})=-\frac{1}{3}\Delta^{\mu\nu}(p_{\mathrm{on}})\mathcal{S}+G^{[\mu\nu]}+\mathcal{T}^{\mu\nu},
\end{equation}
where $\mathcal{S}$, $G^{[\mu\nu]}$ and $\mathcal{T}^{\mu\nu}$
are defined in Eqs. (14) and (15) of Ref. \citep{Dong:2023cng}.

%done, 2025.2.3, 10:50

According to Eq. (\ref{eq:Gless0}), the MVSD can be inversely obtained
from the Green's function as
\begin{align}
f_{\lambda_{1}\lambda_{2}}^{(0)}(x,{\bf p})= & \frac{1}{\pi}\int_{0}^{\infty}dp_{0}p_{0}\epsilon_{\mu}^{*}(\lambda_{1},\mathbf{p})\epsilon_{\nu}(\lambda_{2},\mathbf{p})G_{(0)<}^{\mu\nu}(x,\mathbf{p}),\nonumber \\
\mathrm{Tr}f^{(0)}(x,{\bf p})= & -\frac{1}{\pi}\int_{0}^{\infty}dp_{0}p_{0}\Delta_{\mu\nu}(p_{\mathrm{on}})G_{(0)<}^{\mu\nu}(x,\mathbf{p}).
\end{align}
We assume that above relations hold at any order
\begin{align}
f_{\lambda_{1}\lambda_{2}}(x,{\bf p})= & \frac{1}{\pi}\int_{0}^{\infty}dp_{0}p_{0}\epsilon_{\mu}^{*}(\lambda_{1},p)\epsilon_{\nu}(\lambda_{2},p)G_{<}^{\mu\nu}(x,p),\nonumber \\
\mathrm{Tr}f(x,{\bf p})= & -\frac{1}{\pi}\int_{0}^{\infty}dp_{0}p_{0}\Delta_{\mu\nu}(p)G_{<}^{\mu\nu}(x,p),\label{eq:MVSD-function}
\end{align}
where we have generalized the polarization (field) vector $\epsilon^{\mu}(\lambda,{\bf p})$
to the off-shell four-momentum $p$
\begin{eqnarray}
\epsilon^{\mu}(\lambda,p) & \equiv & \left(\frac{\mathbf{p}\cdot\boldsymbol{\epsilon}_{\lambda}}{\sqrt{p^{2}}},\boldsymbol{\epsilon}_{\lambda}+\frac{\mathbf{p}\cdot\boldsymbol{\epsilon}_{\lambda}}{\sqrt{p^{2}}(p^{0}+\sqrt{p^{2}})}\mathbf{p}\right).
\end{eqnarray}
One can check that $\epsilon^{\mu}(\lambda,p)$ satisfies
\begin{equation}
-\sum_{\lambda}\epsilon^{\mu}(\lambda,p)\epsilon^{\nu*}(\lambda,p)=\Delta^{\mu\nu}(p)\equiv g^{\mu\nu}-\frac{p^{\mu}p^{\nu}}{p^{2}}.
\end{equation}
In this paper we use the off-shell field vector, which is different
from our previous work in Ref. \citep{Dong:2024nxj}, because the
off-shell effect for the $\rho^{0}$ meson is much more significant
than the $\phi$ meson and the expansion in powers of $(p^{0}-E_{p})$
for $\rho^{0}$ does not work well.

In order to calculate the spin density matrix $\rho_{\lambda_{1}\lambda_{2}}$
which is the normalized $f_{\lambda_{1}\lambda_{2}}$, $\rho_{\lambda_{1}\lambda_{2}}\equiv f_{\lambda_{1}\lambda_{2}}/\mathrm{Tr}f$,
from the Wigner function, we define the projector
\begin{equation}
L_{\mu\nu}(\lambda_{1},\lambda_{2},p)\equiv\epsilon_{\mu}^{*}(\lambda_{1},p)\epsilon_{\nu}(\lambda_{2},p)+\frac{1}{3}\Delta_{\mu\nu}(p)\delta_{\lambda_{1}\lambda_{2}}.
\end{equation}
Using Eq. (\ref{eq:MVSD-function}) and the above expression for $L_{\mu\nu}$,
we obtain
\begin{eqnarray}
\int_{0}^{\infty}\frac{dp_{0}}{2\pi}2p_{0}L_{\mu\nu}(\lambda_{1},\lambda_{2},p)G_{<}^{\mu\nu}(x,p) & = & f_{\lambda_{1}\lambda_{2}}(x,\mathbf{p})-\frac{1}{3}\delta_{\lambda_{1}\lambda_{2}}\mathrm{Tr}f(x,\mathbf{p}).
\end{eqnarray}
So the deviation of $\rho_{\lambda_{1}\lambda_{2}}$ from its equilibrium
value (without polarization) $(1/3)\delta_{\lambda_{1}\lambda_{2}}$
can be written as
\begin{align}
\delta\rho_{\lambda_{1}\lambda_{2}}(x,\mathbf{p})\equiv & \rho_{\lambda_{1}\lambda_{2}}(x,\mathbf{p})-\frac{1}{3}\delta_{\lambda_{1}\lambda_{2}}\nonumber \\
= & \frac{\int_{0}^{\infty}dp_{0}p_{0}L_{\mu\nu}(\lambda_{1},\lambda_{2},p)G_{<}^{\mu\nu}(x,p)}{-\int_{0}^{\infty}dp_{0}p_{0}\Delta_{\mu\nu}(p)G_{<}^{\mu\nu}(x,p)}.\label{eq:spin-alignment}
\end{align}
Since $\rho_{00}-1/3$ and $\mathrm{Re}\rho_{-1,1}$ are most relevant
to the $\gamma$ correlator in search for the CME signal \citep{Yin:2024dnu},
we will focus on $\delta\rho_{00}$ and $\mathrm{Re}\delta\rho_{-1,1}$
in this paper.

%done, 2025.2.3, 17:00

In order to calculate $\rho_{\lambda_{1}\lambda_{2}}$ in Eq. (\ref{eq:spin-alignment}),
we will evaluate the Green's function $G_{<}^{\mu\nu}(x,p)$ by expanding
$G_{<}^{\mu\nu}(x,p)$ in powers of external sources: the leading
order (LO) contribution $G_{<,\mathrm{LO}}^{\mu\nu}(x,p)$ without
space-time derivatives and the next-to-leading order (NLO) contribution
$G_{<,\mathrm{NLO}}^{\mu\nu}(x,p)$ with space-time derivatives as
the linear response correction.

%done, 2025.2.4, 6:00

\section{Dyson-Schwinger equation and spectral function \label{sec:Dyson-Schwinger-Equation-and}}

In this section, we will derive the spectral function for the $\rho^{0}$
meson from the Dyson-Schwinger equation for the two-point Green's
function at the leading order. The $\rho\pi\pi$ vertex is taken from
the chiral effective Lagrangian.

%done, 2025.2.4, 7:40

\subsection{Dyson-Schwinger equation on CTP}

We start from the Dyson-Schwinger equation on the CTP for vector mesons
\citep{Sheng:2022ffb,Wagner:2023cct}
\begin{eqnarray}
G^{\mu\nu}(x_{1},x_{2}) & = & G_{(0)}^{\mu\nu}(x_{1},x_{2})+\int_{C}d^{4}x_{1}^{\prime}d^{4}x_{2}^{\prime}G_{(0),\rho}^{\mu}(x_{1},x_{1}^{\prime})\Sigma_{\;\;\sigma}^{\rho}(x_{1}^{\prime},x_{2}^{\prime})G^{\sigma\nu}(x_{2}^{\prime},x_{2}),\label{eq:DSE-1}
\end{eqnarray}
where $\int_{C}$ denotes the integral on the CTP contour, and $\Sigma^{\mu\nu}(x_{1},x_{2})$
is the self-energy. Acting $G_{(0)}^{-1,\mu\nu}$ on both sides of
Eq. (\ref{eq:DSE-1}), we obtain
\begin{eqnarray}
-i\left[g_{\;\rho}^{\mu}(\partial_{x_{1}}^{2}+m_{V}^{2})-\partial_{x_{1}}^{\mu}\partial_{\rho}^{x_{1}}\right]G^{\rho\nu}(x_{1},x_{2}) & = & g^{\mu\nu}\delta_{\mathrm{C}}^{(4)}\left(x_{1}-x_{2}\right)+\int_{C}d^{4}x^{\prime}\Sigma_{\;\;\sigma}^{\mu}(x_{1},x^{\prime})G^{\sigma\nu}(x^{\prime},x_{2}),\label{eq:DSE-2}
\end{eqnarray}
where the delta-function is defined on the CTP
\begin{equation}
\delta_{\text{C}}^{(4)}(x_{1}-x_{2})=\delta^{(3)}({\bf x}_{1}-{\bf x}_{2})\begin{cases}
\delta(x_{1}^{0}-x_{2}^{0}), & x_{1}^{0},x_{2}^{0}\in t_{+}\\
-\delta(x_{1}^{0}-x_{2}^{0}), & x_{1}^{0},x_{2}^{0}\in t_{-}\\
0, & \mathrm{otherwise}
\end{cases}.
\end{equation}
Here $t_{\pm}$ denotes the positive (upper sign) and negative (lower
sign) time branch respectively. Using Eqs. (\ref{eq:Green-function-less},\ref{eq:Green-function-R}),
Eq. (\ref{eq:DSE-2}) can be decomposed into the matrix form
\begin{eqnarray}
 &  & -i\left[g_{\;\rho}^{\mu}(\partial_{x_{1}}^{2}+m_{V}^{2})-\partial_{x_{1}}^{\mu}\partial_{\rho}^{x_{1}}\right]\left(\begin{array}{cc}
0 & G_{A}^{\rho\nu}\\
G_{R}^{\rho\nu} & G_{C}^{\rho\nu}
\end{array}\right)(x_{1},x_{2})\nonumber \\
 & = & \left(\begin{array}{cc}
0 & 1\\
1 & 0
\end{array}\right)g^{\mu\nu}\delta^{(4)}(x_{1}-x_{2})\nonumber \\
 &  & +\int dx^{\prime}\left(\begin{array}{cc}
\Sigma_{A,\rho}^{\mu} & 0\\
\Sigma_{C,\rho}^{\mu} & \Sigma_{R,\rho}^{\mu}
\end{array}\right)(x_{1},x^{\prime})\left(\begin{array}{cc}
0 & G_{A}^{\rho\nu}\\
G_{R}^{\rho\nu} & G_{C}^{\rho\nu}
\end{array}\right)(x^{\prime},x_{2}).
\end{eqnarray}
The Dyson-Schwinger equation for the retarded Green's function $G_{R}$
is
\begin{eqnarray}
-i\left[g_{\;\rho}^{\mu}(\partial_{x_{1}}^{2}+m_{V}^{2})-\partial_{x_{1}}^{\mu}\partial_{\rho}^{x_{1}}\right]G_{R}^{\rho\nu}(x_{1},x_{2}) & = & g^{\mu\nu}\delta^{(4)}(x_{1}-x_{2})+\int dx^{\prime}\Sigma_{R,\rho}^{\mu}(x_{1},x^{\prime})G_{R}^{\rho\nu}(x^{\prime},x_{2}).
\end{eqnarray}
Adopting the Wigner transformation defined in Eq. (\ref{eq:Wigner-tans})
and assuming translation invariance for two-point functions which
is equivalent to neglecting space-time derivative terms (nonlocal
terms), we obtain the equation for $G_{R}$ in momentum space
\begin{equation}
i\left[g_{\;\rho}^{\mu}(p^{2}-m_{V}^{2})-p^{\mu}p_{\rho}\right]G_{R}^{\rho\nu}(p)=g^{\mu\nu}+\Sigma_{R,\rho}^{\mu}(p)G_{R}^{\rho\nu}(p),\label{eq:DSE-R}
\end{equation}
which is the starting point to derive the spectral function for $\rho^{0}$.

%done, 2025.2.4, 10:40

\subsection{Self-energy}

The $\rho\pi$ interaction can be described by the chiral effective
Lagrangian \citep{Gale:1990pn}
\begin{eqnarray}
\mathcal{L} & = & -\frac{1}{4}F^{\mu\nu}F_{\mu\nu}+\frac{1}{2}m_{\rho}^{2}A^{\mu}A_{\mu}+\left|D^{\mu}\phi\right|^{2}-m_{\pi}^{2}\left|\phi\right|^{2},
\end{eqnarray}
where $A^{\mu}$ and $\phi$ denote the field of $\rho^{0}$ and $\pi^{\pm}$
respectively, $F_{\mu\nu}=\partial_{\mu}A_{\nu}-\partial_{\nu}A_{\mu}$
is the field strength tensor for $\rho^{0}$, and $D_{\mu}=\partial_{\mu}-ig_{V}A_{\mu}$
is the covariant derivative. The interaction part of the Lagrangian
is
\begin{equation}
\mathcal{L}_{\mathrm{int}}=ig_{V}A^{\mu}\left(\phi^{\dagger}\partial_{\mu}\phi-\phi\partial_{\mu}\phi^{\dagger}\right)+g_{V}^{2}A_{\mu}A^{\mu}\phi^{\dagger}\phi,\label{eq:Lagrangian-int}
\end{equation}
where the first and second terms give the $\rho\pi\pi$ and $\rho\rho\pi\pi$
vertices respectively.

%done, 2025.2.4, 10:40

\begin{figure}
\includegraphics[scale=0.5]{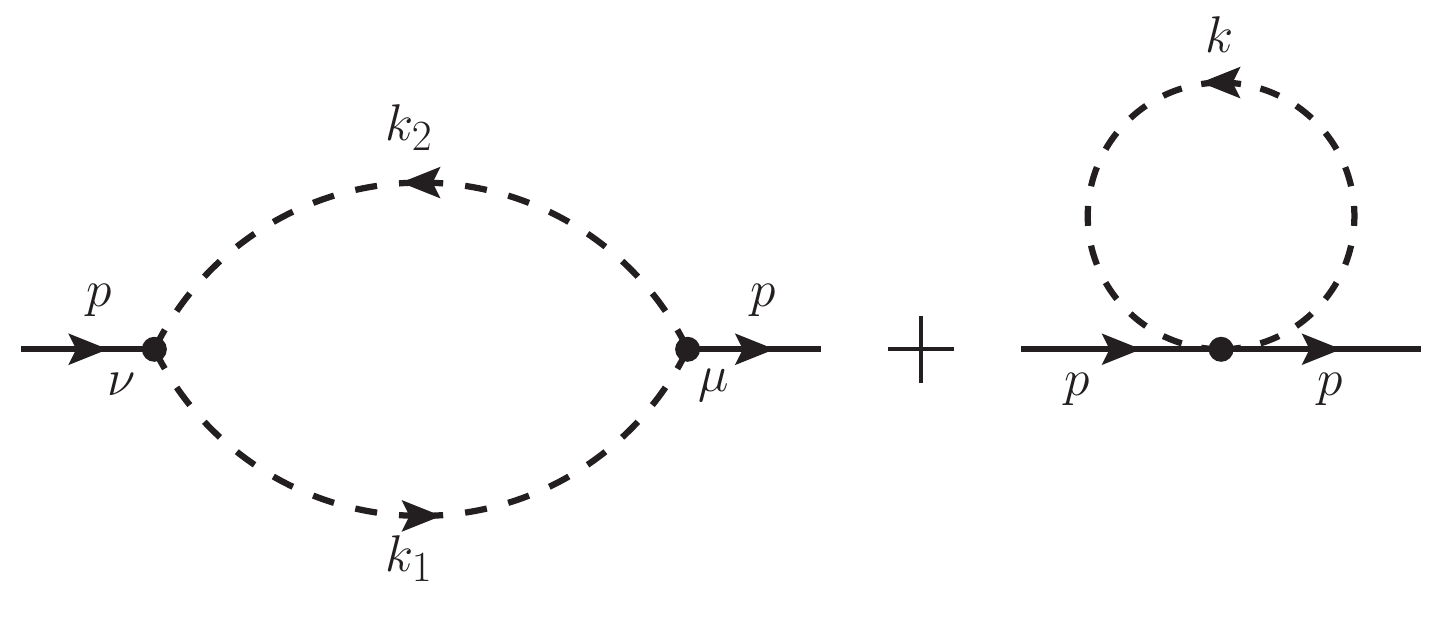}

\caption{The Feynman diagrams for the self-energy term of $\rho^{0}$ meson
with momentum $p$. The solid line and dashed line denote the propagators
of $\rho^{0}$ and $\pi^{\pm}$ meson respectively, and the arrow
on the $\pi$ propagator denotes the momentum direction of $\pi^{+}$.
\label{fig:Sedlf-energy}}
\end{figure}

The Feynman diagrams for the self-energy are shown in Fig. (\ref{fig:Sedlf-energy}).
The retarded self-energy is
\begin{eqnarray}
\Sigma_{R}^{\mu\nu}(p) & = & \Sigma_{F}^{\mu\nu}(p)-\Sigma_{<}^{\mu\nu}(p)\nonumber \\
 & = & -g_{V}^{2}\int\frac{d^{4}k_{1}}{(2\pi)^{4}}\int\frac{d^{4}k_{2}}{(2\pi)^{4}}(2\pi)^{4}\delta^{(4)}\left(p-k_{1}+k_{2}\right)\left(k_{1}^{\mu}+k_{2}^{\mu}\right)\left(k_{1}^{\nu}+k_{2}^{\nu}\right)S^{F}(k_{1})S^{F}(k_{2})\nonumber \\
 &  & +g_{V}^{2}\int\frac{d^{4}k_{1}}{(2\pi)^{4}}\int\frac{d^{4}k_{2}}{(2\pi)^{4}}(2\pi)^{4}\delta^{(4)}\left(p-k_{1}+k_{2}\right)\left(k_{1}^{\mu}+k_{2}^{\mu}\right)\left(k_{1}^{\nu}+k_{2}^{\nu}\right)S^{<}(k_{1})S^{>}(k_{2})\nonumber \\
 &  & +2g_{V}^{2}ig^{\mu\nu}\int\frac{d^{4}k}{(2\pi)^{4}}\frac{i}{k^{2}-m_{\pi}^{2}+i\epsilon}\nonumber \\
 & = & I_{\mathrm{vac}}^{\mu\nu}(p)+I_{\mathrm{med}}^{\mu\nu}(p),
\end{eqnarray}
where $I_{\mathrm{vac}}^{\mu\nu}(p)$ and $I_{\mathrm{med}}^{\mu\nu}(p)$
denote the vacuum and medium parts respectively. They can be put into
the forms \citep{Gale:1990pn},
\begin{eqnarray}
I_{\mathrm{vac}}^{\mu\nu}(p) & = & i\Delta^{\mu\nu}(p)\Pi_{\mathrm{vac}}(p),\nonumber \\
I_{\mathrm{med}}^{\mu\nu}(p) & = & i\Delta_{L}^{\mu\nu}(p)\Pi_{L,\mathrm{med}}(p)+i\Delta_{T}^{\mu\nu}(p)\Pi_{T,\mathrm{med}}(p),\label{eq:SE-vac-med}
\end{eqnarray}
where the transverse and longitudinal projectors are defined as
\begin{eqnarray}
\Delta_{L}^{\mu\nu}(p) & \equiv & \frac{\Delta^{\mu\rho}u_{\rho}\Delta^{\nu\sigma}u_{\sigma}}{\Delta^{\rho\sigma}u_{\rho}u_{\sigma}},\nonumber \\
\Delta_{T}^{\mu\nu}(p) & \equiv & \Delta^{\mu\nu}-\Delta_{L}^{\mu\nu}.
\end{eqnarray}
Here $u^{\mu}$ is the flow velocity, in the comoving frame, it becomes
$u^{\mu}=u_{\mu}=(1,\mathbf{0})$. In Eq. (\ref{eq:SE-vac-med}),
$\Pi_{\mathrm{vac}}$, $\Pi_{L,\mathrm{med}}$ and $\Pi_{T,\mathrm{med}}$
are given as
\begin{eqnarray}
\Pi_{\mathrm{vac}} & = & \frac{g_{V}^{2}}{(4\pi)^{2}}p^{2}\int_{0}^{1}dx(2x-1)^{2}\left[\log\left|\frac{m_{\pi}^{2}-x(1-x)p^{2}}{m_{\pi}^{2}-x(1-x)m_{\rho}^{2}}\right|-i\pi\theta\left(x(1-x)p^{2}-m_{\pi}^{2}\right)\right],\\
\Pi_{L,\mathrm{med}}(p) & = & \frac{g_{V}^{2}}{8\pi^{2}}\frac{p^{2}}{|\mathbf{p}|^{2}}\left[-4J_{0}(-1,2)+\frac{2}{|\mathbf{p}|}\left(J_{+}(p,1,1)+J_{-}(p,1,1)\right)\right.\nonumber \\
 &  & \left.+\frac{2p_{0}}{|\mathbf{p}|}\left(J_{+}(p,0,1)-J_{-}(p,0,1)\right)+\frac{p_{0}^{2}}{2|\mathbf{p}|}\left(J_{+}(p,-1,1)+J_{-}(p,-1,1)\right)\right],\\
\Pi_{T,\mathrm{med}}(p) & = & \frac{g_{V}^{2}}{8\pi^{2}}\left[\frac{2\left(p_{0}^{2}+|\mathbf{p}|^{2}\right)}{|\mathbf{p}|^{2}}J_{0}(-1,2)+\frac{1}{|\mathbf{p}|}\left(J_{+}(p,-1,3)+J_{-}(p,-1,3)\right)\right.\nonumber \\
 &  & -\frac{\left(p^{2}\right)^{2}}{4|\mathbf{p}|^{3}}\left(J_{+}(p,-1,1)+J_{-}(p,-1,1)\right)-\frac{p^{2}p_{0}}{|\mathbf{p}|^{3}}\left(J_{+}(p,0,1)-J_{-}(p,0,1)\right)\nonumber \\
 &  & \left.-\frac{p_{0}^{2}}{|\mathbf{p}|^{3}}\left(J_{+}(p,1,1)+J_{-}(p,1,1)\right)\right],
\end{eqnarray}
where momentum functions $J_{\pm}(p,n_{1},n_{2})$ and $J_{0}(n_{1},n_{2})$
are defined as
\begin{eqnarray}
J_{+}(p,n_{1},n_{2}) & \equiv & \int_{0}^{\infty}d|\mathbf{k}|E_{k}^{n_{1}}|\mathbf{k}|^{n_{2}}\left[f_{\pi^{+}}(E_{k})+f_{\pi^{-}}(E_{k})\right]\nonumber \\
 &  & \times\ln\frac{p^{2}+2p_{0}E_{k}+2|\mathbf{p}||\mathbf{k}|+i\epsilon}{p^{2}+2p_{0}E_{k}-2|\mathbf{p}||\mathbf{k}|+i\epsilon},\label{eq:J+}\\
J_{-}(p,n_{1},n_{2}) & \equiv & \int_{0}^{\infty}d|\mathbf{k}|E_{k}^{n_{1}}|\mathbf{k}|^{n_{2}}\left[f_{\pi^{+}}(E_{k})+f_{\pi^{-}}(E_{k})\right]\nonumber \\
 &  & \times\ln\frac{p^{2}-2p_{0}E_{k}+2|\mathbf{p}||\mathbf{k}|+i\epsilon}{p^{2}-2p_{0}E_{k}-2|\mathbf{p}||\mathbf{k}|+i\epsilon},\label{eq:J-}\\
J_{0}(n_{1},n_{2}) & \equiv & \int_{0}^{\infty}d|\mathbf{k}|E_{k}^{n_{1}}|\mathbf{k}|^{n_{2}}\left[f_{\pi^{+}}(E_{k})+f_{\pi^{-}}(E_{k})\right].\label{eq:J0}
\end{eqnarray}
In Eqs. (\ref{eq:J+}-\ref{eq:J0}), We have assumed that $\pi^{\pm}$
are in thermal equilibrium, so $f_{\pi^{+}}$ and $f_{\pi^{-}}$ are
Bose-Einstein distribution depending on the temperature $T$ and chemical
potential $\pm\mu_{\pi}$. Here we have written the self-energy in
the same forms as in our previous work \citep{Dong:2024nxj}, and
one can prove that it is identical to the results in Ref. \citep{Gale:1990pn}.

%done, 2025.2.4, 15:00

Finally, the total self-energy can be written as
\begin{eqnarray}
\Sigma_{R}^{\mu\nu} & = & i\Delta_{L}^{\mu\nu}\Pi_{L}+i\Delta_{T}^{\mu\nu}\Pi_{T},\label{eq:SE}
\end{eqnarray}
where $\Pi_{L}$ and $\Pi_{T}$ are defined as
\begin{eqnarray}
\Pi_{L} & \equiv & \Pi_{\mathrm{vac}}+\Pi_{L,\mathrm{med}},\nonumber \\
\Pi_{T} & \equiv & \Pi_{\mathrm{vac}}+\Pi_{T,\mathrm{med}}.
\end{eqnarray}

Note that Eqs. (\ref{eq:J+}-\ref{eq:J0}) are results in the flow's
comoving frame. To obtain their expressions in the lab frame we have
to make a Lorentz boost. In this paper, we work in the flow's comoving
frame.

%done, 2025.2.4, 15:20

\subsection{Spectral function}

Inserting Eq. (\ref{eq:SE}) into Eq. (\ref{eq:DSE-R}), we can solve
$G_{R}^{\mu\nu}(p)$ as a function of $\Pi_{L}$ and $\Pi_{T}$,
\begin{equation}
G_{R}^{\mu\nu}(p)=\frac{-i}{p^{2}-m_{V}^{2}-\Pi_{L}}\Delta_{L}^{\mu\nu}+\frac{-i}{p^{2}-m_{V}^{2}-\Pi_{T}}\Delta_{T}^{\mu\nu}+\frac{i}{m_{V}^{2}}\frac{p^{\mu}p^{\nu}}{p^{2}},\label{eq:G_R}
\end{equation}
where we have used
\begin{align}
\Delta_{L,\rho}^{\mu}\Delta_{L}^{\rho\nu} & =\Delta_{L}^{\mu\nu},\;\;\;\Delta_{T,\rho}^{\mu}\Delta_{T}^{\rho\nu}=\Delta_{T}^{\mu\nu},\nonumber \\
\Delta_{L,\rho}^{\mu}\Delta_{T}^{\rho\nu} & =p_{\mu}\Delta_{L}^{\mu\nu}=p_{\mu}\Delta_{T}^{\mu\nu}=0.
\end{align}
We can read out the longitudinal and transverse spectral function
from Eq. (\ref{eq:G_R}) as
\begin{eqnarray}
\rho_{L}(p) & = & -\mathrm{Im}\frac{1}{p^{2}-m_{V}^{2}-\Pi_{L}},\nonumber \\
\rho_{T}(p) & = & -\mathrm{Im}\frac{1}{p^{2}-m_{V}^{2}-\Pi_{T}}.
\end{eqnarray}
From the relation $\tilde{G}_{<}^{\mu\nu}(p)=2in_{B}(p_{0})\mathrm{Im}\tilde{G}_{R}^{\mu\nu}(p)$
\citep{fetter2003quantum,zubarev1997statistical} ($G_{i}^{\mu\nu}=i\tilde{G}_{i}^{\mu\nu}$
for $i=<,>,F,\bar{F},G,A$), the leading order Wigner function reads
\begin{equation}
G_{<,\mathrm{LO}}^{\mu\nu}(p)=-2n_{B}(p^{0})\left[\Delta_{L}^{\mu\nu}\rho_{L}(p)+\Delta_{T}^{\mu\nu}\rho_{T}(p)\right].\label{eq:Gless-LO}
\end{equation}
Here $n_{B}(p_{0})$ is the Bose-Einstein distribution. Inserting
Eq. (\ref{eq:Gless-LO}) into Eq. (\ref{eq:spin-alignment}), we can
obtain the leading order contribution to the spin density matrix.

\section{Linear Response Theory \label{sec:Linear-Response-Theory}}

In this section, we will derive the NLO correction to the Wigner function
by linear response theory. We will use the Kubo formula derived in
Zubarev's approach \citep{Zubarev_1979,Hosoya:1983id,Becattini:2019dxo}.

The linear response of an observable $\left\langle \hat{O}(x)\right\rangle $
for the operator $\hat{O}(x)$ to the perturbation $\partial_{\mu}\beta_{\nu}$
reads \citep{Becattini:2019dxo}
\begin{eqnarray}
\left\langle \hat{O}(x)\right\rangle -\left\langle \hat{O}(x)\right\rangle _{\mathrm{LE}} & = & \partial_{\mu}\beta_{\nu}(x)\lim_{K^{\mu}\rightarrow0}\frac{\partial}{\partial K_{0}}\nonumber \\
 &  & \times\mathrm{Im}\left[iT(x)\int_{-\infty}^{t}d^{4}x^{\prime}\left\langle \left[\hat{O}(x),T^{\mu\nu}(x^{\prime})\right]\right\rangle _{\mathrm{LE}}e^{-iK\cdot(x^{\prime}-x)}\right],\label{eq:Kubo-formulae}
\end{eqnarray}
where $\left\langle \hat{O}(x)\right\rangle \equiv\mathrm{Tr}\left[\hat{\rho}\hat{O}(x)\right]$
is the expectation value in non-equilibrium while $\left\langle \hat{O}(x)\right\rangle _{\mathrm{LE}}\equiv\mathrm{Tr}\left[\hat{\rho}_{\mathrm{LE}}\hat{O}(x)\right]$
is the expectation value in local equilibrium, $\hat{\rho}$ and $\hat{\rho}_{\mathrm{LE}}$
are non-equilibrium and local equilibrium density operators respectively,
$\beta^{\mu}(x)\equiv u^{\mu}(x)/T(x)$ with $u^{\mu}(x)$ and $T(x)$
being the flow velocity and temperature respectively, and $T^{\mu\nu}$
is the energy-momentum tensor for the vector field give by
\begin{eqnarray}
T^{\mu\nu} & = & F_{\;\rho}^{\mu}F^{\rho\nu}+m_{V}^{2}A^{\mu}A^{\nu}-g^{\mu\nu}\left(-\frac{1}{4}F_{\rho\sigma}F^{\rho\sigma}+\frac{1}{2}m_{V}^{2}A_{\rho}A^{\rho}\right).
\end{eqnarray}

%done, 2025.2.5, 10:20

We choose the operator $\hat{O}(x)$ in Eq. (\ref{eq:Kubo-formulae})
to be the Wigner function operator
\begin{eqnarray}
\hat{G}_{<}^{\mu\nu}(x,p) & \equiv & \int d^{4}ye^{ip\cdot y}A^{\nu}\left(x-\frac{1}{2}y\right)A^{\mu}\left(x+\frac{1}{2}y\right).
\end{eqnarray}
The correction to $G_{<}^{\mu\nu}(x,p)$ from the perturbation $\partial_{\mu}\beta_{\nu}$
can be written as
\begin{eqnarray}
G_{<,\mathrm{NLO}}^{\mu\nu}(x,p) & = & 2T\xi_{\gamma\lambda}\frac{\partial n_{B}(p_{0})}{\partial p_{0}}I^{\mu\nu\gamma\lambda}(p),\label{eq:Gless-NLO}
\end{eqnarray}
where $\xi_{\gamma\lambda}=\partial_{(\gamma}\beta_{\lambda)}$ denotes
the thermal shear stress tensor, and the tensor $I^{\mu\nu\gamma\lambda}(p)$
is defined as
\begin{eqnarray}
I^{\mu\nu\gamma\lambda}(p) & \equiv & -\left[g^{\lambda\gamma}\left(p^{2}-m^{2}\right)-2p^{\lambda}p^{\gamma}\right]\left(\Delta_{L}^{\mu\nu}\rho_{L}^{2}+\Delta_{T}^{\mu\nu}\rho_{T}^{2}\right)\nonumber \\
 &  & +2\left(p^{2}-m^{2}\right)\left(\Delta_{L}^{\mu\lambda}\rho_{L}+\Delta_{T}^{\mu\lambda}\rho_{T}\right)\left(\Delta_{L}^{\nu\gamma}\rho_{L}+\Delta_{T}^{\nu\gamma}\rho_{T}\right).\label{eq:I-term}
\end{eqnarray}
The detailed derivation of $I^{\mu\nu\gamma\lambda}(p)$ can be found
in \citep{Li:2022vmb,Dong:2024nxj}.

%done, 2025.2.5, 10:30

Now the total Wigner function is
\begin{eqnarray}
G_{<}^{\mu\nu}(x,p) & = & G_{<,\mathrm{LO}}^{\mu\nu}(x,p)+G_{<,\mathrm{NLO}}^{\mu\nu}(x,p),\label{eq:Gless}
\end{eqnarray}
where the LO and NLO contributions are given by Eqs. (\ref{eq:Gless-LO})
and (\ref{eq:Gless-NLO}) respectively. Inserting Eq. (\ref{eq:Gless})
into Eq. (\ref{eq:spin-alignment}), the deviation of the spin density
matrix from 1/3 can be expressed as
\begin{eqnarray}
\delta\rho_{\lambda_{1}\lambda_{2}}(x,\mathbf{p}) & = & \frac{\int_{0}^{\infty}dp_{0}p_{0}L_{\mu\nu}(\lambda_{1},\lambda_{2},p)\left[G_{<,\mathrm{LO}}^{\mu\nu}(x,p)+G_{<,\mathrm{NLO}}^{\mu\nu}(x,p)\right]}{-\int_{0}^{\infty}dp_{0}p_{0}\Delta_{\mu\nu}(p)\left[G_{<,\mathrm{LO}}^{\mu\nu}(x,p)+G_{<,\mathrm{NLO}}^{\mu\nu}(x,p)\right]}.\label{eq:spin-alignment-result}
\end{eqnarray}
In order to numerically calculate the spin density matrix that can
be compared with experimental data in momentum space, we have to integrate
over $x$ on the freeze-out hypersurface for both numerator and denominator
in Eq. (\ref{eq:spin-alignment-result}) as
\begin{eqnarray}
\delta\rho_{\lambda_{1}\lambda_{2}}(\mathbf{p}) & = & \frac{\int_{0}^{\infty}dp_{0}\int d\Sigma^{\mu}p_{\mu}L_{\mu\nu}(\lambda_{1},\lambda_{2},p)\left[G_{<,\mathrm{LO}}^{\mu\nu}(x,p)+G_{<,\mathrm{NLO}}^{\mu\nu}(x,p)\right]}{-\int_{0}^{\infty}dp_{0}\int d\Sigma^{\mu}p_{\mu}\Delta_{\mu\nu}(p)\left[G_{<,\mathrm{LO}}^{\mu\nu}(x,p)+G_{<,\mathrm{NLO}}^{\mu\nu}(x,p)\right]},\label{eq:spin-alignment-momentum}
\end{eqnarray}
where $\Sigma^{\mu}$ is the freeze-out hypersurface. If the system
evolves homogeneously in time, $\Sigma^{0}$ is actually the space
volume, so $d\Sigma^{0}=d^{3}x$. The above formula is the starting
point for the numerical calculation in the next section.

%done, 2025.2.5, 10:30

\section{Numerical Results \label{sec:Numerical-Result}}

In this section, we will numerically calculate $\rho_{00}$ and $\mathrm{Re}\rho_{-1,1}$
by Eq. (\ref{eq:spin-alignment-momentum}). The parameters are chosen
as $m_{\rho}=$770 MeV, $m_{\pi}=$139 MeV, $g_{V}=$6.07 for the
$\rho\pi\pi$ coupling constant \citep{Gale:1990pn}, $T=$150 MeV
for the freeze-out temperature. The rapidity range is set to $|Y|<1$.
The freeze-out hypersurface and thermal shear stress tensor are calculated
by hydrodynamical model CLVisc at $\sqrt{s_{NN}}=$11.5, 19.6, 27,
39, 62.4 and 200 GeV and 20-50\% centrality \citep{Pang:2018zzo,Wu:2021fjf,Wu:2022mkr}.
Here we assume that the thermal shear stress tensor does not change
a lot before and after hadronization, so we can apply it to the $\rho\pi$
system.

%done, 2025.2.5, 11:30

\subsection{Results for $\rho_{00}$}

From Eq. (\ref{eq:spin-alignment-momentum}), the LO contribution
to $\rho_{00}$ involves
\begin{eqnarray}
L_{\mu\nu}(0,0,p)G_{<,\mathrm{LO}}^{\mu\nu}(x,p) & = & -2\left(\frac{p_{y}^{2}}{|\mathbf{p}|^{2}}-\frac{1}{3}\right)n_{B}(p_{0})\left[\rho_{T}(p)-\rho_{L}(p)\right],\nonumber \\
-\Delta_{\mu\nu}(p)G_{<,\mathrm{LO}}^{\mu\nu}(x,p) & = & 2n_{B}(p_{0})\left[2\rho_{T}(p)+\rho_{L}(p)\right].\label{eq:LO-spin-alignment}
\end{eqnarray}
We see that the spin alignment as a function of the momentum may be
non-vanishing at the LO although the momentum integrated spin alignment
at the same order should be zero. The NLO contribution involves
\begin{eqnarray}
L_{\mu\nu}(0,0,p)G_{<,\mathrm{NLO}}^{\mu\nu}(x,p) & = & -2T\xi_{\lambda\gamma}\frac{\partial n_{B}(p_{0})}{\partial p_{0}}\nonumber \\
 &  & \times\left\{ \left[g^{\lambda\gamma}\left(p^{2}-m^{2}\right)-2p^{\lambda}p^{\gamma}\right]L_{\mu\nu}(0,0,p)\left(\Delta_{L}^{\mu\nu}\rho_{L}^{2}+\Delta_{T}^{\mu\nu}\rho_{T}^{2}\right)\right.\nonumber \\
 &  & \left.-2\left(p^{2}-m^{2}\right)L_{\mu\nu}(0,0,p)\left(\Delta_{L}^{\mu\lambda}\rho_{L}+\Delta_{T}^{\mu\lambda}\rho_{T}\right)\left(\Delta_{L}^{\nu\gamma}\rho_{L}+\Delta_{T}^{\nu\gamma}\rho_{T}\right)\right\} ,\label{eq:NL-num}\\
-\Delta_{\mu\nu}(p)G_{<,\mathrm{NLO}}^{\mu\nu}(x,p) & = & 2T\xi_{\lambda\gamma}\frac{\partial n_{B}(p_{0})}{\partial p_{0}}\left\{ \left[g^{\lambda\gamma}\left(p^{2}-m^{2}\right)-2p^{\lambda}p^{\gamma}\right]\left(2\rho_{T}^{2}+\rho_{L}^{2}\right)\right.\nonumber \\
 &  & \left.-2\left(p^{2}-m^{2}\right)\left(\Delta_{L}^{\lambda\gamma}\rho_{L}^{2}+\Delta_{T}^{\lambda\gamma}\rho_{T}^{2}\right)\right\} .\label{eq:NL-den}
\end{eqnarray}
The NLO contribution is proportional to the thermal shear stress tensor
$\xi_{\lambda\gamma}$, so it may contribute to the momentum integrated
spin alignment. It is important to note that the coefficient of $\xi_{\lambda\gamma}$
increases with increasing $p$, so the expansion will break down for
large $p$. In other words, the linear response theory only works
for low momenta.

%done, 2025.2.5, 12:30; corrected 2025.3.12

\begin{figure}
\subfloat[\label{fig:a}]{\includegraphics[scale=0.5]{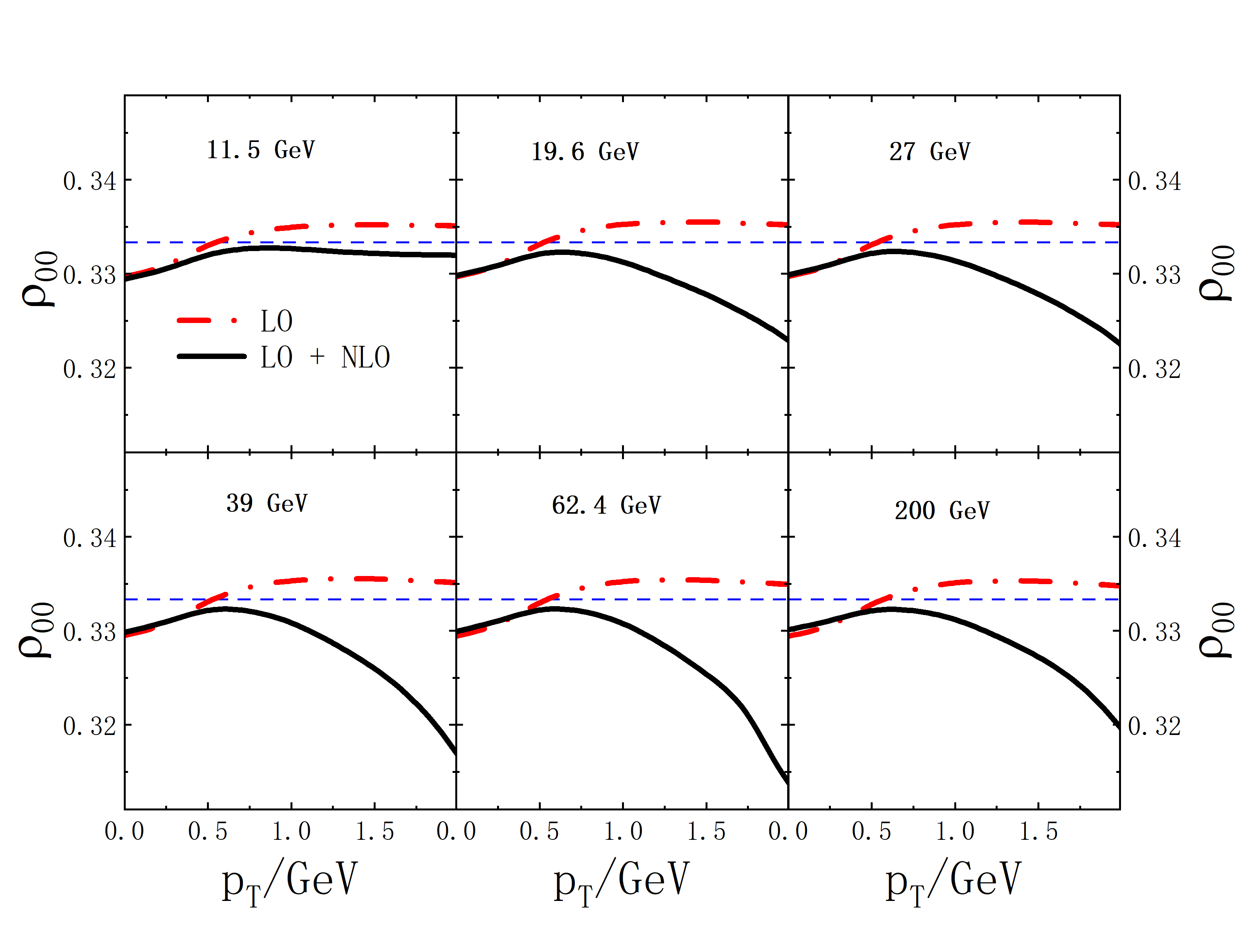}

}

\subfloat[\label{fig:b}]{\includegraphics[scale=0.5]{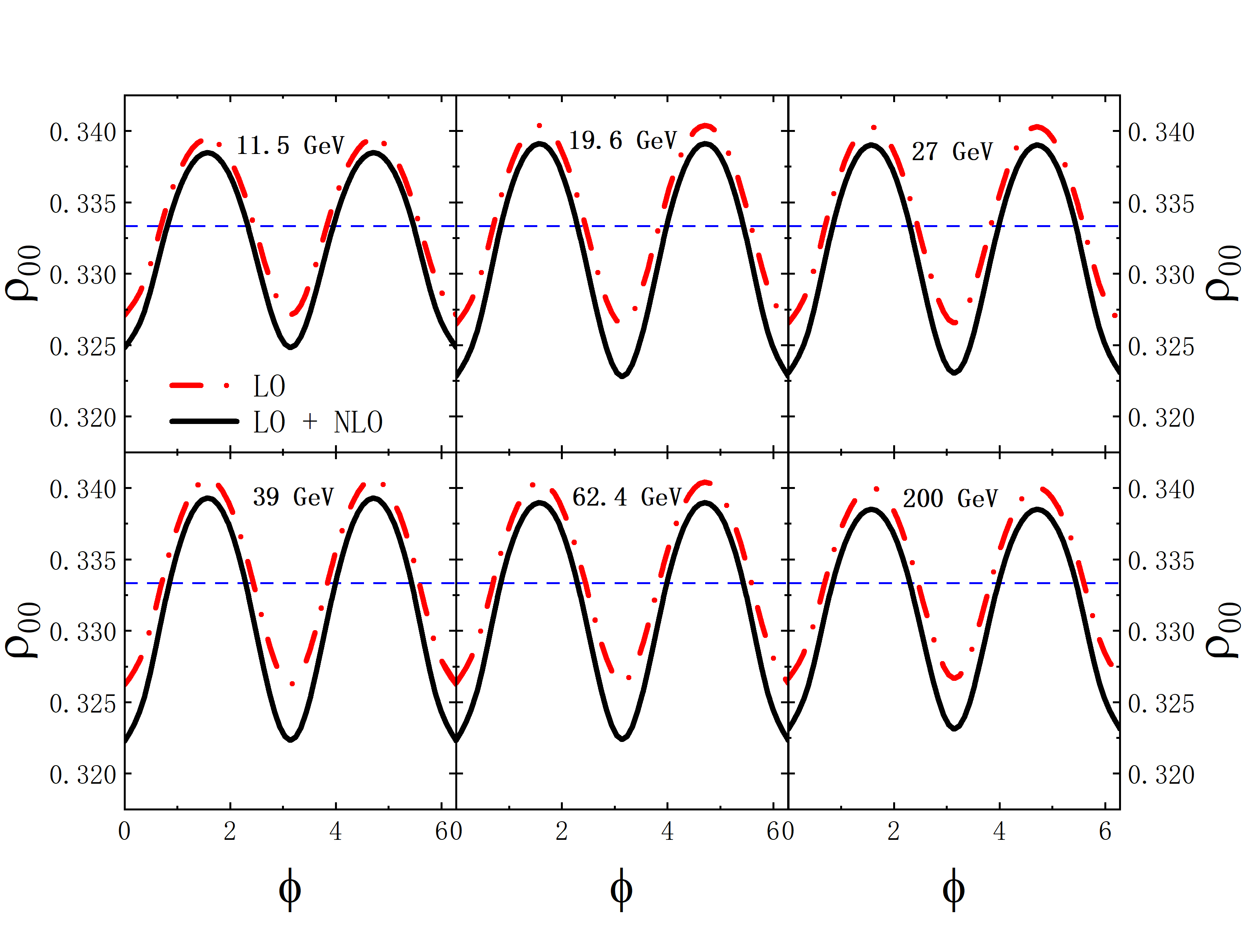}

}

\caption{The spin alignment of $\rho^{0}$ mesons at different collision energies
as functions of (a) transverse momentum $p_{T}$ and (b) azimuthal
angle $\phi$ in the central rapidity region $|Y|<1$. The red dash-dotted
lines are the LO contribution, while the black solid lines are the
sum over the LO and NLO contribution. \label{fig:The-spin-alignment}}

\end{figure}

\begin{figure}
\includegraphics[scale=0.3]{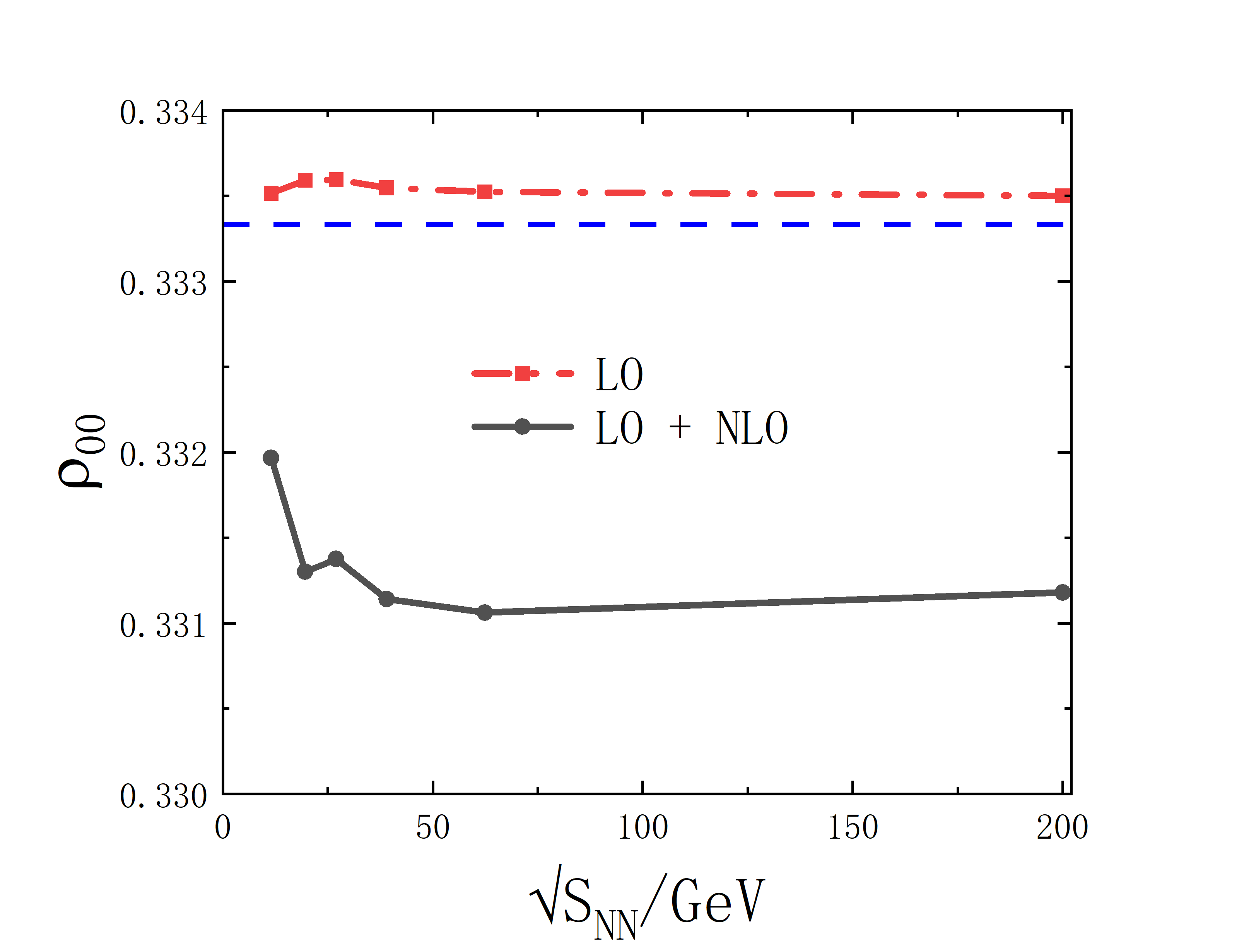}\caption{The spin alignment of $\rho^{0}$ mesons in the central rapidity region
$|Y|<1$ as functions of the collision energy. \label{fig:The-spin-alignment-total}}

\end{figure}

The spin alignment as functions of the transverse momentum $p_{T}$
and azimuthal angle $\phi$ in the central rapidity $|Y|<1$ is shown
in Fig. (\ref{fig:The-spin-alignment}), and $\rho_{00}$ as functions
of the collision energy is shown in Fig. (\ref{fig:The-spin-alignment-total}).
We can see in Fig. (\ref{fig:a}) that $\delta\rho_{00}$ (the spin
alignment) at the LO is negative for $p_{T}\lesssim$0.5 GeV and positive
for $p_{T}\apprge$0.5 GeV but with very small magnitude of the order
$10^{-3}$, while its NLO contribution is negative and its magnitude
increases with $p_{T}$ due to the term proportional to $\xi_{\lambda\gamma}p^{\lambda}p^{\gamma}$
in Eqs. (\ref{eq:NL-num}-\ref{eq:NL-den}). It turns out that the
total values of $\delta\rho_{00}$ from the LO and NLO are negative
in the region up to $p_{T}=$2 GeV. Note that the non-vanishing values
of the LO contribution to $\delta\rho_{00}$ are due to the space-time
integration on the freezeout hypersurface and the rapidity range.
The azimuthal angle $\phi$ dependence of $\delta\rho_{00}$ is shown
in Fig. (\ref{fig:b}) in an oscillation pattern with the magnitude
at the LO being about $10^{-2}$. We see that $\delta\rho_{00}$ is
positive for $\phi=\pi/2,3\pi/2$ and negative for $\phi=0,\pi$.
The NLO contribution to $\delta\rho_{00}$ is slightly smaller than
zero since it is suppressed by $n_{B}(p_{0})$ for high $p_{T}$ in
$p_{T}$ integration. Fig. (\ref{fig:The-spin-alignment-total}) shows
that the LO contribution to $\delta\rho_{00}$ is independent of the
collision energy, while the NLO contribution is negative and decreases
with the collision energy when it is lower than about 50 GeV.

%done, 2025.2.5, 13:00; corrected 2025.3.12

%done, 2025.2.5, 13:00

\subsection{Results for $\mathrm{Re}\rho_{-1,1}$}

From Eq. (\ref{eq:spin-alignment-momentum}), the LO and NLO contributions
to $\mathrm{Re}\rho_{-1,1}$ involve
\begin{eqnarray}
\mathrm{Re}L_{\mu\nu}(-1,1,p)G_{<,\mathrm{LO}}^{\mu\nu}(x,p) & = & -n_{B}(p_{0})\frac{p_{x}^{2}-p_{z}^{2}}{|\mathbf{p}|^{2}}\left[\rho_{T}(p)-\rho_{L}(p)\right],\label{eq:LO-rho-11}\\
\mathrm{Re}L_{\mu\nu}(-1,1,p)G_{<,\mathrm{NLO}}^{\mu\nu}(x,p) & = & -2T\xi_{\lambda\gamma}\frac{\partial n_{B}(p_{0})}{\partial p_{0}}\left\{ \left[g^{\lambda\gamma}\left(p^{2}-m^{2}\right)-2p^{\lambda}p^{\gamma}\right]\right.\nonumber \\
 &  & \times\mathrm{Re}L_{\mu\nu}(-1,1,p)\left(\Delta_{L}^{\mu\nu}\rho_{L}^{2}+\Delta_{T}^{\mu\nu}\rho_{T}^{2}\right)\nonumber \\
 &  & -2\left(p^{2}-m^{2}\right)\mathrm{Re}L_{\mu\nu}(-1,1,p)\nonumber \\
 &  & \left.\times\left(\Delta_{L}^{\mu\lambda}\rho_{L}+\Delta_{T}^{\mu\lambda}\rho_{T}\right)\left(\Delta_{L}^{\nu\gamma}\rho_{L}+\Delta_{T}^{\nu\gamma}\rho_{T}\right)\right\} ,\label{eq:NLO-rho-11}
\end{eqnarray}
Here the denominator is the same as in $\rho_{00}$. Similarly, the
LO contribution to $\mathrm{Re}\rho_{-1,1}$ as a momentum function
may be non-vanishing although the momentum integrated one should be
vanishing. The NLO contribution may give rise to both integrated and
un-integrated non-zero $\mathrm{Re}\rho_{-1,1}$.

%done, 2025.2.5, 13:30

\begin{figure}
\subfloat[\label{fig:a-2}]{\includegraphics[scale=0.5]{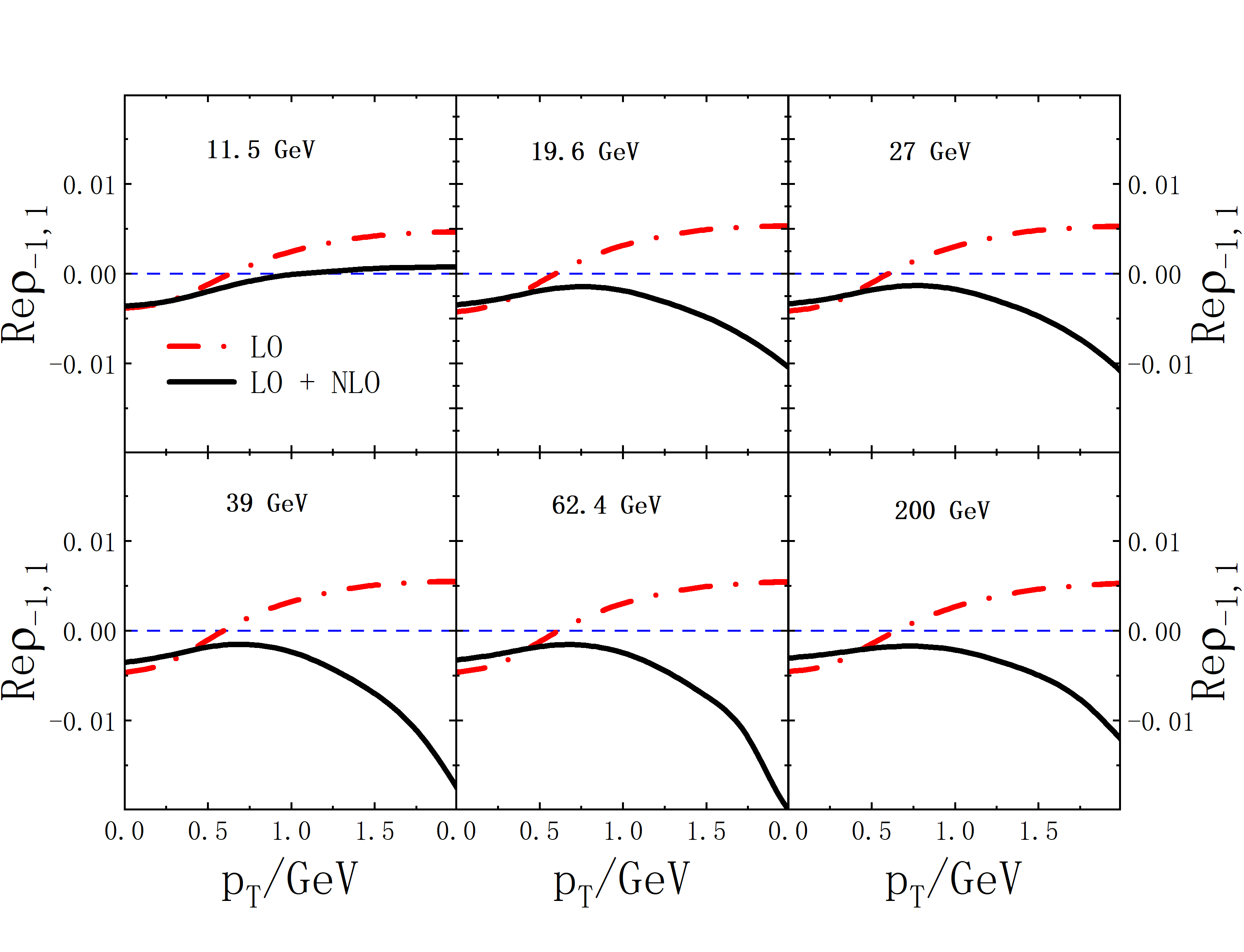}

}

\subfloat[\label{fig:b-2}]{\includegraphics[scale=0.5]{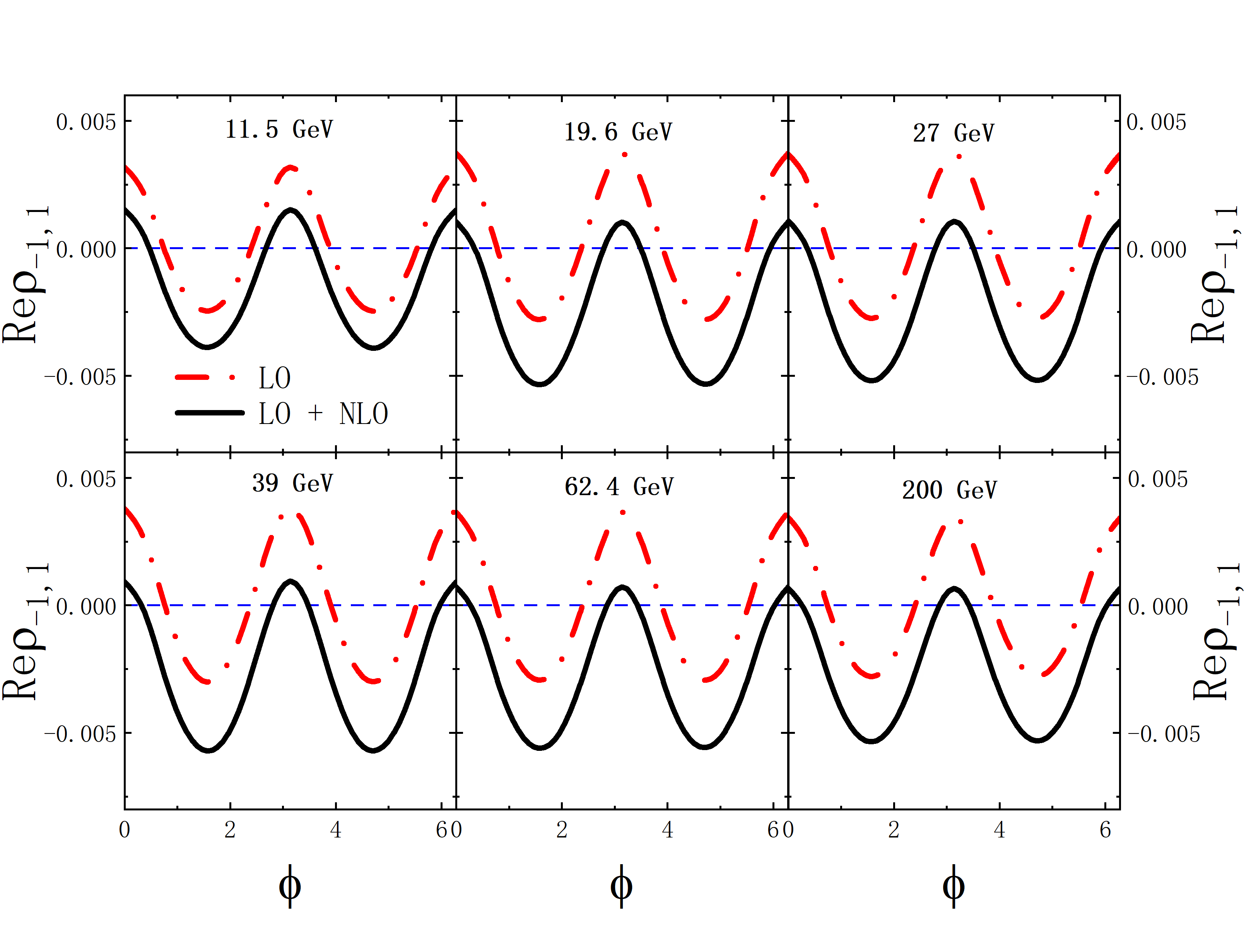}

}

\caption{The off-diagonal element $\mathrm{Re}\rho_{-1,1}$ of $\rho^{0}$
mesons at different collision energies as functions of (a) transverse
momentum $p_{T}$ and (b) azimuthal angle $\phi$ in the central rapidity
region $|Y|<1$. The red dash-dotted lines are the LO contribution,
while the black solid lines are the sum over the LO and NLO contribution.
\label{fig:The-Rerho-11}}

\end{figure}

\begin{figure}
\includegraphics[scale=0.3]{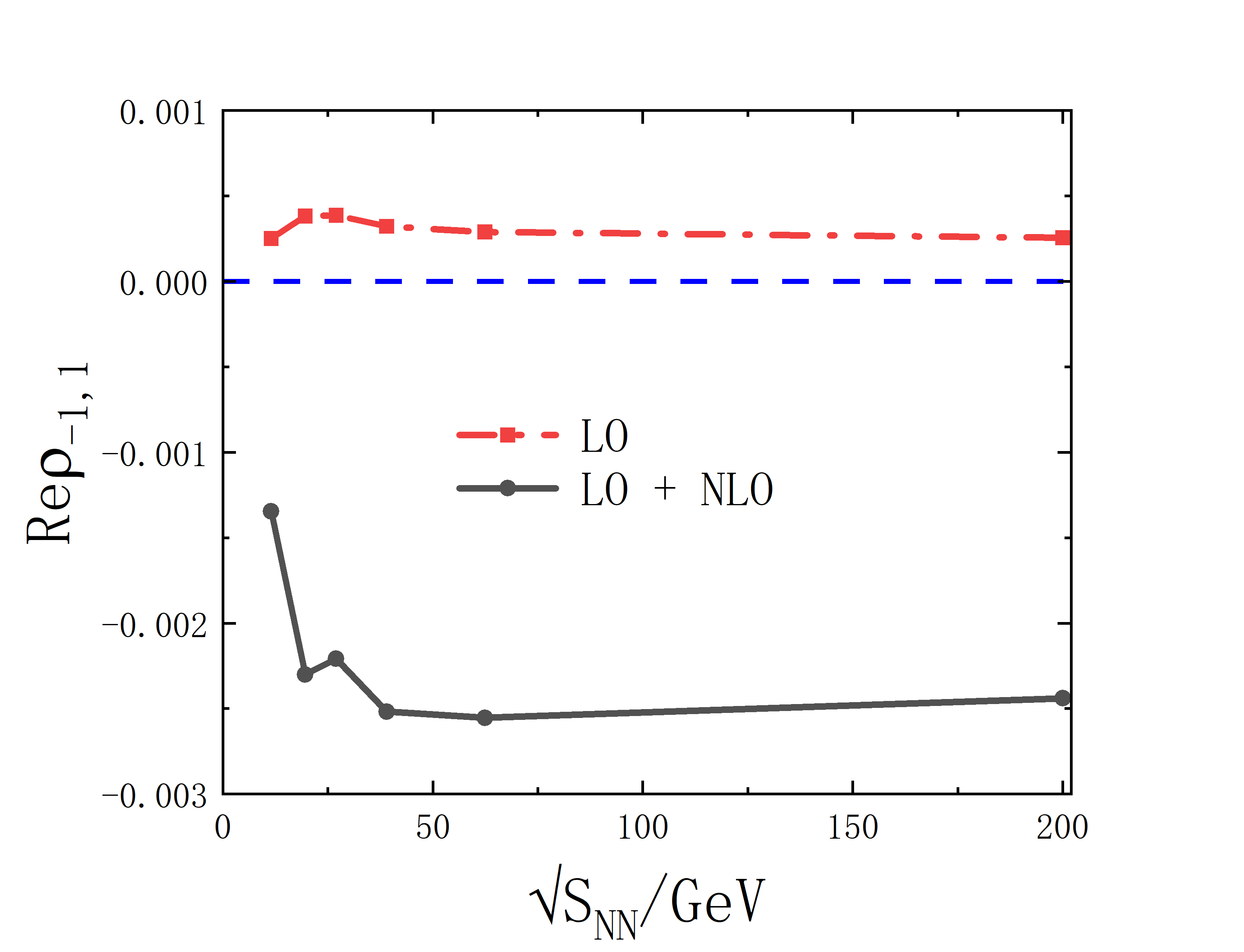}

\caption{The off-diagonal element $\mathrm{Re}\rho_{-1,1}$ of $\rho^{0}$
mesons in the central rapidity region $|Y|<1$ as functions of the
collision energy. \label{fig:The-Rerho-11-total}}

\end{figure}

The off-diagonal element $\mathrm{Re}\rho_{-1,1}$ as functions of
$p_{T}$ and $\phi$ in the central rapidity range $|Y|<1$ is presented
in Fig. (\ref{fig:The-Rerho-11}), and its dependence on the collision
energy is presented in Fig. (\ref{fig:The-Rerho-11-total}). In Fig.
(\ref{fig:a-2}), we can see that the LO contribution is negative
for low $p_{T}$ and positive for high $p_{T}$ with the magnitude
of the order $10^{-3}$. The NLO contribution is negative and its
magnitude increases with $p_{T}$, which is also a result of the $\xi_{\lambda\gamma}p^{\lambda}p^{\gamma}$
term. As shown in Fig. (\ref{fig:b-2}), $\mathrm{Re}\rho_{-1,1}$
as functions of $\phi$ has an oscillation pattern similar to $\delta\rho_{00}$
but with an opposite sign, due to the different signs of $p_{x}^{2}$
terms in Eqs. (\ref{eq:LO-spin-alignment}) and (\ref{eq:LO-rho-11})
(we can express $p_{y}^{2}$ as $|\mathbf{p}|^{2}-p_{x}^{2}-p_{z}^{2}$).
The collision energy dependence of $\mathrm{Re}\rho_{-1,1}$ also
has the similar behavior to $\delta\rho_{00}$, as shown in Fig. (\ref{fig:The-Rerho-11-total}).

%done, 2025.2.5, 13:50; corrected 2025.3.12, 8:20

%done, 2025.2.5, 13:50

In conclusion, the contributions from the spectral function and from
the thermal shear stress tensor to the spin density matrix of $\rho^{0}$
mesons are of the order of $10^{-3}\sim10^{-2}$. For un-integrated
quantities, $\delta\rho_{00}(p_{T})$ and $\mathrm{Re}\rho_{-1,1}(p_{T})$
are all negative and decrease with $p_{T}$, and $\delta\rho_{00}(\phi)$
and $\mathrm{Re}\rho_{-1,1}(\phi)$ show oscillation patterns but
with opposite signs. The momentum integrated $\delta\rho_{00}$ and
$\mathrm{Re}\rho_{-1,1}$ are all negative and decrease with the collision
energy when the collision energy is less than about 50 GeV, while
they are almost constant when the collision energy is higher.

%done, 2025.2.5, 14:00; corrected 2025.3.12

\section{Summary and conclusions \label{sec:Conclusion}}

We study the spin density matrix of neutral $\rho$ mesons contributed
from the spectral function and thermal shear tensor with the Kubo
formula in the linear response theory. We introduce the two-point
Green's function in the CTP formalism which defines the Wigner function
and the MVSD in phase space for neutral $\rho$ mesons. As the leading
order contribution, the spectral function of the neutral $\rho$ meson
can be obtained from the Dyson-Schwinger equation for the retarded
Green's function with $\rho\pi\pi$ and $\rho\rho\pi\pi$ interactions
in a thermal pion gas. Then the next-to-leading order contribution
from the thermal shear tensor can be calculated through the Kubo formula
in the linear response theory. Finally we present numerical results
for $\rho_{00}$ and $\mathrm{Re}\rho_{-1,1}$ which have a dominant
effect on the $\gamma$ correlator in search for the CME. The numerical
results for the thermal shear tensor are needed for evaluating the
spin density matrix, which can be calculated by hydrodynamical models.
The numerical results give negative values for $\delta\rho_{00}$
and $\mathrm{Re}\rho_{-1,1}$ as functions of the transverse momentum
at the order $10^{-2}$, while their values as functions of the azimuthal
angle show oscillation patterns. The momentum integrated $\delta\rho_{00}$
and $\mathrm{Re}\rho_{-1,1}$ are all negative and decrease with the
collision energy when the collision energy is less than about 50 GeV,
while they are almost constant for higher collision energies.

%done, 2025.2.6, 12:00; corrected 2025.3.12
\begin{acknowledgments}
The work is supported in part by the National Natural Science Foundation
of China (NSFC) under Grant No. 12135011.
\end{acknowledgments}

\bibliographystyle{h-physrev}
\phantomsection\addcontentsline{toc}{section}{\refname}\bibliography{ref-linear-rho-spin}

\end{document}